\documentstyle[aps,epsfig,twocolumn]{revtex}
\newcommand{\eeq}{\end{equation}}
\newcommand{\beq}{\begin{equation}}
\newcommand{\nuq}[1]{\label{#1} \eeq}

\begin{document}
\title{
Regular and Anomalous Quantum Diffusion in the
Fibonacci Kicked Rotator.
}
\author{G. Casati$^{(1)}$, G. Mantica$^{(1)}$ and D.L. Shepelyansky$^{(2)}$ \\
$^{(1)}$Int. Ctr for the Study of Dynamical Systems, Universit\`a
dell'Insubria,  Via Valleggio 11 I-22100 Como Italy \\
 and Istituto Nazionale Fisica della Materia, unit\`a di  Milano\\
 and Istituto Nazionale Fisica Nucleare, sez. di  Milano\\
$^{(2)}$ Laboratoire de Physique Quantique, UMR 5626 du CNRS,\\ 
Universit\'e Paul Sabatier, F-31062 Toulouse Cedex 4, France } \maketitle

\begin{abstract}
We study the dynamics of a quantum rotator kicked 
according to the almost-periodic Fibonacci sequence.
A special numerical technique allows us to carry on this
investigation for as many as 
$10^{12}$ kicks. It is shown that above a critical
kick strength the excitation of the system is well described
by  regular diffusion, while
below this border it becomes anomalous, and sub-diffusive.
A law for the dependence of the exponent of anomalous 
sub-diffusion on the system parameters is established numerically.
The analogy between these results and
quantum diffusion in models of quasi-crystal and in the kicked Harper system
is discussed.
\end{abstract}
\pacs{PACS numbers 05.45.Mt 02.30.Tb}

\section{Introduction}

It can be reasonably argued that the
quantum kicked rotator \cite{rotator}
is the oldest, and more thoroughly investigated
system in the realm of the so-called quantum chaos.
Yet, we pretend in this paper that recourse to this system is
still justified, and can provide new dynamical phenomena worth of investigation.
Consider in fact the problem of {\em transport} in quantum
extended systems: for instance, a wave-packet is initially
localized on a few states of a quantum lattice, and is then left
free to evolve. A deep analysis carried out on model systems
\cite{simo1,simo2,mahi,suto,hirabe,abehir} with
nearest neighbours interactions has shown that this dynamics is
either {\em localized}--this is typically the case of random
couplings, or, in the case of almost periodic couplings, {\em
anomalous} ({\em i.e.} the diffusion coefficient is either zero or
infinity--almost never finite) and {\em intermittent}
\cite{pirl,phd1,palad,size,ig1,igm,piech}.

It is well known that the kicked rotator for typical values of the parameters
is akin to the first example just presented: its quantum dynamics is
(eventually) localized
in the action lattice. Therefore, it seems to feature a far simpler
dynamics than the second class of systems mentioned above, or than
its not-too-close relative, the {\em quantum kicked Harper} model
\cite{karp}, in which anomalous diffusion 
\cite{diffu} has been documented.
A way to cure this simplicity is offered by an idea 
profitably employed in spin and few-levels systems \cite{spin0,spinz,spin}:
rather than acting upon the system with a periodic impulsive force,
we let the sequence of time-intervals between different "kicks"
be almost-periodic. In few-level systems, this idea has
lead to interesting spectral properties and dynamical behaviours.

In applying this procedure to the kicked rotator we hope so to combine
the complexity of an almost periodic interval sequence and the
richness of the infinite Hilbert space of the unperturbed problem, while
maintaining at the same time a sufficient simplicity for the theoretical
analysis:
for this, we elect to investigate the effect of a Fibonacci sequence
of unitary evolution operators.

This paper presents the results of a numerical analysis
of the system so defined.
In previous studies of almost-periodic kicking of the quantum rotator,
a linear increase of the energy has been observed \cite{combe,cesar},
which amounts to  {\em regular} diffusion.
In alternative, power-law increase of energy with a non-unit exponent
defines {\em anomalous} diffusion. We shall attach special importance to
this exponent, which describes the {\em growth} of the
number of the excited states of the system.
Energy is just one of the indicators of the motion of
this quantum system. We shall use here a larger set of indicators
to describe, in the long time,
the growth of the excitation of the system. 
Each of these quantities will be characterized by a 
power-law asymptotic behaviour, which in turn will define an associated
{\em growth exponent}.

This analysis will show that two dynamical regions are
present in this system. For large kicking strength
the motion is diffusive: the growth exponents are equal to one half,
and a proper diffusion coefficient can be defined.
To the contrary, when the kick strength is less than
a certain threshold,
the motion typically features {\em anomalous sub-diffusion}:
growth exponents still exist, but are smaller than one half.
The investigation is further carried on by varying the kick strength:
in so doing, a power-law dependence of the growth exponents is observed.
It is interesting to note that a similar situation 
has been found in quasi-crystals in two dimensions
\cite{sire}: in these systems, for strong hopping amplitude the 
wave-packet spreading over a
two-dimensional lattice is diffusive, while, in the opposite case,
spreading is characterized by anomalous sub-diffusion. 

This paper is organized as follows: in the next section we
define the model. In Sect. III we introduce growth indicators
and their power-law exponents
to gauge anomalous diffusion in the
dynamics of an initially localized wave-packet.
The numerical algorithm and the numerically computed 
long time dynamics of the
system are presented in Sect. IV. 
Finally, we allow a random element
in the definition of the dynamical system to verify the robustness
of the obtained results, and to support
the conjecture that anomalous diffusion in this model is induced by
the almost-periodicity of the sequence of unitary operators.

\section{The Dynamical System}

The basis of our investigation is the quantum kicked rotator 
\cite{rotator}: this is the unitary evolution generated in the usual Hilbert space
$L_2(S_{2 \pi})$ by the operators
 \beq
    U(k,T) = e^{-i k \cos \theta }
   \;   e^{-i \frac{T}{2}  {\hat n}^2}
\nuq{unit1} which depend on the two parameters $k,T$ and where the momentum
operator is $\hat n = -i
\partial_\theta$. The
corresponding classical system is the well known Chirikov standard
map \cite{stmap}, which is obtained letting $k$ grow, and $T$ go to zero,
while keeping their product fixed. This product defines the
classical chaos parameter, $K=kT$. In the classical case the system
becomes globally diffusive for $K>0.9716...$.

Now, it is well known that for large values of $K$, and typical
irrational values of $\frac{T}{2 \pi}$, classical and quantum dynamics of the
kicked rotator are profoundly different: this is the content of the
quantum phenomenon of dynamical localization, which has been 
well described in the literature, and which predicts that the quantum
evolution of an initially focused wave-function is almost
periodic in time, and localized on the momentum space 
of the unperturbed system at $k=0$ \cite{rotator}.

In this paper, we study a variation on this theme: while in the
usual approach the unitary evolution of eq. (\ref{unit1}) is acted
repeatedly, at {\em equally spaced} times $t_n = nT$, in our model
the length of the time intervals in between kicks
will {\em not} be constant. Equivalently, we can say that 
the operators $U$ act now in Fibonacci sequence
to produce the full quantum evolution $\cal U$:
 \beq
  {\cal U} (t_n) = U(k,\tau_n) \circ
   U(k,\tau_{n-1}) \circ \ldots \circ
   U(k,\tau_{1}) ,
\nuq{unit2}
where $t_n = \sum_{j=1}^n \tau_j$ is time, 
and where $\circ$ denotes operator composition. As is immediately
observed, operators in eq. (\ref{unit2}) are characterised by the
same value of the quantum kick amplitude $k$, and  differ solely
by the value of $T$. The sequence of values of this latter is
determined by letting $\tau_j$ to
take either one of two possible values, which for convenience we
denote by $A$ and $B$. We choose for
$A$ and $B$ the values
 \[
A = \frac{2 \pi}{\lambda}, \;\; B = \frac{2 \pi}{\lambda^2} ,
\]
where 
$\lambda=1.3247...$
is an algebraic number, that solves the
equation $\lambda^3 -  \lambda - 1 = 0$. 
In a sense, $A,B$ is
the {\em most irrational} couple of numbers.

We line up  $A$'s and $B$'s according to the well known Fibonacci
sequence: this is obtained by an accretion rule on the initial
finite pieces of the infinite sequence, $\sigma_1=A$ and
$\sigma_2=A,B$. According to this rule, the next finite piece is
obtained by joining these two sequences, the first at the end of
the second: $\sigma_3=A,B,A$. In general,
 \beq
   \sigma_{n+1} = \sigma_{n} \cup \sigma_{n-1} ,
\nuq{cat2}
 where the symbol $\cup$ means here joining the right
and left sequences {\em in the specified order}, so that
$\sigma_4=A,B,A,A,B$, and so on. It is then easy to see that the
length of $\sigma_n$ is equal to the $n$-th Fibonacci number,
$F_n$. Eq. (\ref{cat2}) expresses a property that will be
instrumental in the following to obtain the quantum evolution 
up to large times.

\section{Growth Exponents of an Initially Localized Wave-function}

We now let $\psi(0)$ be the initial state of the evolution, and we
compute the time-dynamics $ \psi(t_n) =  {\cal U}(t_n) \psi(0)$.
Typically, we choose for initial state $\psi(0)$ the usual $e_0$
momentum basis state of $L_2(S_{2 \pi})$, the full basis being
$e_n=e^{i n \theta}$, with $n \in {\bf Z}$.

As we have already remarked, the quantum kicked
rotator, for {\em typical} irrational values
of the kicking period $T$, displays  the phenomenon of {\em quantum
localization}: action in the system eventually saturates in time, and
quantum diffusion stops. On the other hand, 
for kick sequences with {\em random} 
time intervals after each kick, the increase in momentum is diffusive and
never comes to an end \cite{italo}. Of particular
interest is then to study the kicking sequence defined in the
previous section, which is neither periodic, nor random: it
belongs to the family of {\em almost-periodic} sequences.
We shall see that
under these circumstances
dynamical localization
leaves places to a delocalization which shares many
characteristics  with that taking place in almost-periodic
lattice systems.

To study the dynamics of the quantum wave-function, it is
convenient to introduce a number of quantities suitable to define its
spreading: we shall call them collectively {\em growth indicators}.
The first obvious choice are moments.
Let $\nu_\alpha(t)$ be the moment of index
$\alpha$ of the probability distribution over the lattice $\{e_n\}$
given by $|(e_n,\psi(t))|^2$, where obviously $(\cdot,\cdot)$
denotes the scalar product in $L_2(S_{2 \pi})$:
  \beq
  \nu_\alpha(t) :=   \sum_n |(e_n,\psi(t))|^2 |n|^\alpha.
 \nuq{mom1}
For instance, $\nu_2$ is the usual second moment--the dispersion--which,
in the case of regular diffusion, grows {\em linearly} in time.
In general, we cannot expect $\nu_2$ to behave as $t$,
but {\em super-diffusive} and {\em sub-diffusive} growth will be the rule.
For this, we also define the {\em growth exponent} function
$\beta$ by the asymptotic relation
 \beq
    \nu_\alpha(t) \sim t^{\alpha \beta(\alpha)} \;\; \mbox{for} \;\;
  t \rightarrow \infty,
 \nuq{espo1}
 where the asymptotic behaviour is to be understood in
a suitable sense. The function $\beta(\alpha)$ is the so-called
{\em quantum intermittency function}, described
at length in  \cite{pirl,phd1,palad}.

When $\alpha$ tends to zero, the moment $\nu_\alpha$ tends to the
constant $\nu_0=1$. Nonetheless, one can define the limit of the
function $\beta$ for $\alpha$ tending to zero, $\beta(0)$,
a number which also appears in the scaling of the
logarithmic moment:
 \beq
 \sum_{n \neq 0} |(e_n,\psi(t))|^2 \log( |n| ) \sim \beta(0) \log (t) .
 \nuq{logmom}

In the quantities (\ref{espo1})
and (\ref{logmom} the {\em ordering} of the basis is
relevant: a function of $n$ appears in the summations. We can
nonetheless define and compute quantities which do not make
reference to this characteristics.  The first of these is the {\em
entropy}:
 \beq
   - \sum_n |(e_n,\psi(t))|^2 \log (|(e_n,\psi(t))|^2) \sim \gamma +
   \beta_e \log (t),
 \nuq{ent1}
whose asymptotic behaviour (here and in the following to be
understood in the infinite time limit) defines the exponent
$\beta_e$. In the case of localization, $\beta_e$ is null, and
$e^\gamma$ (the constant contribution in eq. (\ref{ent1})) gives a
measure of the localization length. Another basis independent
quantity is the {\em inverse participation ratio},
 \beq
  \sum_n  |(e_n,\psi(t))|^4 \sim t^{-\beta_p},
 \nuq{ipr1}
with its exponent $\beta_p$.

\begin{figure}[htbp]
\centering\epsfig{file=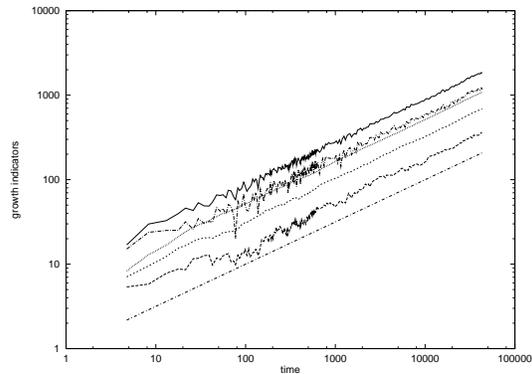,width=0.6\linewidth,angle=270}
\caption{ Regular quantum diffusion. Growth indicators are plotted
versus time given in number
of kicks (in abscissa) for the case $k$ = 10. The lowest curve
(dots-dashes)
is the reference line $\sqrt t$. The other curves are (from bottom
to top): the exponential of the entropy (long dashes), the square root of the
second moment (short dashes), 
the sixth root of the sixth moment (dots), one over the
inverse participation ratio (dots-dashes), and the exponential of the
logarithmic moment (continuous line).} \protect\label{momen}
\end{figure}

We notice that, in principle, the exponents $\beta$ of all the
quantities so far defined are different.  Yet, when finite, they
are all consistent with the fact that the spreading of the wave-packet over
the basis takes place in a power-law fashion.
This is indeed the case: in fig. (\ref{momen}) we  plot the
$\frac{1}{\alpha}$ power of a set of moments $\nu_\alpha$ versus
time, in doubly logarithmic scale, together with one over the
inverse participation ratio, and with the exponential  of the
entropy and of the logarithmic moment. It is immediately
appreciated that quantum diffusion in this case is regular: these
curves are parallel to the reference line $\sqrt t$, so that all
growth exponents are equal to one half.

\begin{figure}[htbp]
\centering\epsfig{file=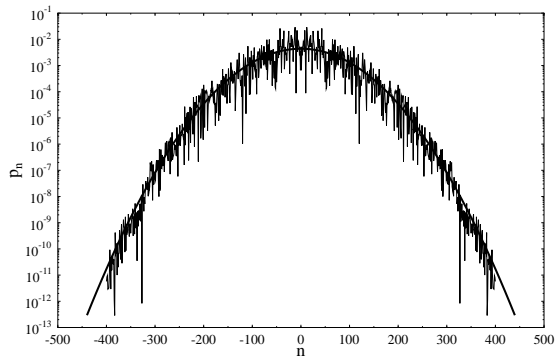,width=0.8\linewidth}
\caption{Wave packet for the case of fig. 1. Plotted are occupation
probabilities $p_n = |(e_n,\psi(t))|^2 $ 
versus site number $n$ at time $t=1.797 \; 10^{4}$ (thin line). 
The thick line is an interpolating normal distribution.}

\protect\label{pacch}
\end{figure}

To render even more evident the fact that we are in the presence of
regular diffusion we plot in fig. (\ref{pacch})
a snapshot of  the wave-packet
at a large time, for the same case of fig. (\ref{momen}):
The fitting line is a normal distribution.

\section{Detecting anomalous diffusion in large time dynamics}

Let us now investigate the effect of varying the
kick strength $k$: in particular, let us decrease it from the
value $k=10$ at which we observed regular diffusion.
Two difficulties arise when performing this analysis:
since we observe a {\em reduced} delocalization of the wave-packet,
very large time-scale computations are required to say something
reliable about the asymptotic behaviour. Secondly, large
{\em systematic} oscillations in the behaviour of the
moments (and of the other quantities) tend to 
overwhelm and submerge the
power-law increase, the more the slower this increase.

Let us first present the cure to the second difficulty: we can
introduce a {\em magnetic flux} $\alpha$ in the evolution
(\ref{unit1}), which is nothing else than replacing the operator 
$\hat n$ with $\hat n + \phi$: $\phi$
is a real number, so that for $\phi = 0 $ we obtain the usual
kicked system. By performing the evolution with a set of values of
$\phi$, and by averaging the results with respect to this sample,
we can reduce the systematic fluctuations super-imposed to the
leading power-law behaviour. Please observe that the classical
dynamics is left invariant by the introduction of this flux.

For the first difficulty, a more complex strategy is required. We
can push the evolution to extremely large times by resorting to a
matrix algorithm. Let ${\cal U}^{F_n}$ be the unitary evolution operator
acting from time zero up to the time of the kick number $F_n$, the
$n$-th Fibonacci number. Then, eq. (\ref{cat2}) implies that these
operators are simply related by
 \beq
   {\cal U}^{F_{n+1}} = {\cal U}^{F_{n-1}} \circ {\cal U}^{F_{n}} ,
\nuq{circ1}
 where $\circ$ is the usual operator multiply, and
where the order of operators is essential, as it was in eq.
(\ref{cat2}). This relation must be initialised by setting
 \beq
\begin{minipage}{5.0cm}
$
   {\cal U}^{F_1} =  U(k,A), $ \\
$   {\cal U}^{F_2} =  U(k,B) \circ U(k,A), $
\end{minipage}
\nuq{circ2} 
so to produce the full sequence
 $\{{\cal U}^{F_n}\}$. In fact, the matrix elements
of $U(K,\tau)$ are explicitly known, and eq. (\ref{circ1})
involves then only a matrix-to-matrix multiplication. Of course, the
{\em cost} of this operation scales as the cube of the size of the
matrix, and this is certainly more expensive than the usual
evolution effected via a sequence of fast Fourier transforms.
Nonetheless, this technique becomes advantageous when dealing with
extremely large times, for in $n$ steps one reaches up to $F_n$
map iterations, and it is well known that Fibonacci numbers
are geometrically increasing in $n$.

We therefore effect the matrix multiplication for $n=1,\ldots,N$
and then look at the zeroth coloumn of the resulting operator
${\cal U}^{F_N}$: this is the evoluted of the initial state $e_0$.
In our numerical experiments, the initial state of the 
evolution will always be $e_0$.
Suitable combinations of the columns can provide the evoluted of
any arbitrary initial state, if desired.

A question of numerical concern is to be discussed prior to
showing the results of these calculations: when dealing with
equations (\ref{circ1},\ref{circ2}) a finite truncation of the
operators involved is necessary. Now, 
for any practical purpose 
$U(k,\tau)$ is 
a banded matrix, and we must obviously choose
the size of the numerical truncation much larger than the band
size. Yet this is not enough: appropriate boundary conditions at
the basis edges must be imposed. Two choices are at hand:
Dirichlet, and Neumann. In the latter case, we are producing the
unitary evolution on the {\em torus}, while in
the first case  we end up with a non-unitary
evolution.

Having in mind what we want to obtain ({\em i.e.} the evolution of
the $e_0$ basis state in the full cylinder space)
we have adopted Dirichlet conditions, for two reasons. First,
when the truncation size is sufficiently large, 
both Dirichlet and Neumann conditions {\em must} produce an
excellent approximation of the true ${\cal U}^{F_N} e_0$. Secondly,
we can gauge numerically two quantities: the normalization of the
zeroth column of ${\cal U}^{F_N}$ (which is a common technique) and
its {\em effective dimension}--which is less common, but more instructive.
In fact, this latter was defined above as the number of components
of ${\cal U}^{F_N} e_0$ of magnitude larger than a certain threshold: this
quantifies the dimension of the wave-packet after $F_N$ kicks,
and can be profitably used to control the optimal size of the
truncation. It must be added that the effective dimension in
itself has physical and mathematical relevance
\cite{size}.

In summary, we have been able to reach easily about sixty
iterations of the Fibonacci multiplication, eq. (\ref{circ2}), which
correspond to more than $10^{12}$ usual iterations, and this with a
basis size of the order of the thousand. Numerical stability, controlled
by various techniques, is here the limiting factor.

\begin{figure}[htbp]
\centering\epsfig{file=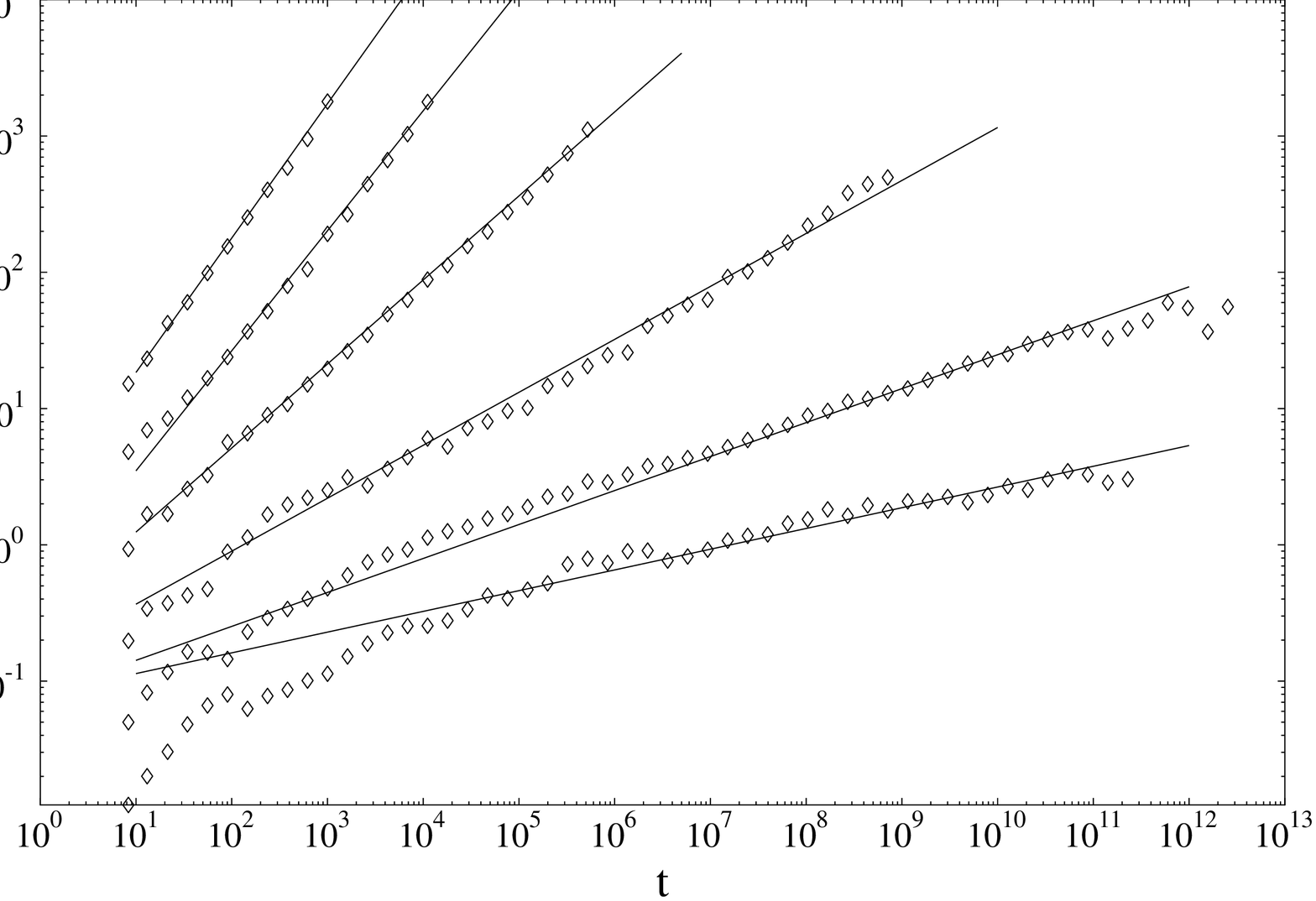,width=0.8\linewidth}
\caption{Second moment $\nu_2$ as a function of time, for
$k=0.125, 0.25, 0.5, 1, 2, 4.$ (bottom to top 
diamond curves). Also shown are
the fitting lines from which the exponent $\beta(2)$ is extracted. }
\protect\label{nu2}
\end{figure}

A set of typical results is shown
in fig. (\ref{nu2}), where we plot the second moment,
$\nu_2$, as a function of time, in doubly logarithmic scale,
for a set of different values of $k$.
We clearly see that these values decrease with the latter, but, more important,
the slopes of the curves ({\em i.e.} the growth exponent $\beta(2)$) also
decrease with $k$.
As a consequence, we can safely conclude that quantum diffusion
is anomalous in this model for a wide set of parameters \cite{infto}.

At this point,
a word must be spent about intermittency: this is present when the function
$\beta(\alpha)$ (see eq. (\ref{espo1})) is not constant. In the case at hand,
the variation of $\beta$ with $\alpha$ is smaller than the
uncertainty with which $\beta$ is determined, both due to numerical
effects, and to the super-imposed oscillations of the moments $\nu_\alpha$.
We can therefore only conclude that intermittency, if present, is low.

\begin{figure}[htbp]
\centering\epsfig{file=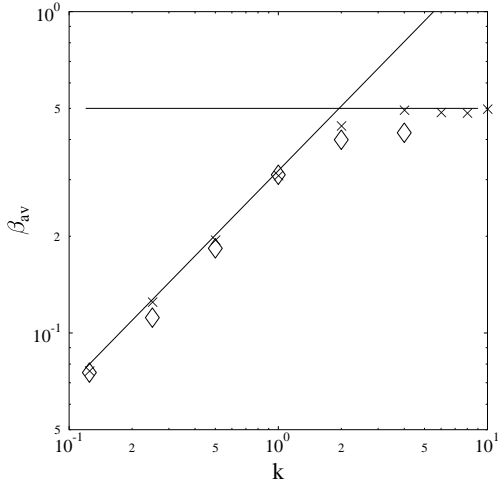,width=1.3\linewidth,angle=0}
\caption{ Growth exponent $\beta_{av}$ (crosses) obtained as the
average of $\beta(\alpha)$, for $\alpha = 2,4,6$ , versus the kick
amplitude $k$. The horizontal line marks the value
$\beta_{av}=\frac{1}{2}$; the fitting line for small values of $k$
grows proportional to $k^\eta$, with $\eta=\frac{2}{3}$. Also reported
(diamonds) are the result obtained when random rotation phases 
replace the irrational phases $Tn^2/2$ (see text).
 }
\protect\label{betak}
\end{figure}

This fact is not totally negative: by averaging over $\alpha$, we can define
a more reliable growth exponent ${\beta}_{av}$ which is now function only
of the kick amplitude $k$. This immediately prompts for the study of this
dependence.
Fig. (\ref{betak}) plots the numerical results obtained
by the procedure exposed above.
Two regions clearly emerge from the investigation of this picture:
for large values of $k$ we observe the regular diffusive value
${\beta}_{av} = \frac{1}{2}$.
Anomalous diffusion is observed for smaller values of $k$:
quite interestingly, we find a power-law behaviour of the
growth exponent of the form 
\beq
  {\beta}_{av} (k) \sim k^\eta,
\nuq{law}
 with $\eta$ very
close to the value $\frac{2}{3}$. The transition from 
anomalous to regular diffusion takes place for $k>k_c$:
for the set of values we have chosen, this critical value $k_c$
is approximately equal to two.

Let us now investigate the origin of anomalous diffusion
in this model.
When acting on the basis set $e_n=e^{i n \theta}$ the free
evolution $e^{i \frac{T}{2} \partial_\theta^2}$ produces the phase factor
$e^{-i \frac{T}{2} n^2}$. The arithmetic nature of $T/2 \pi$ is at the root
of the spectral properties of the usual kicked rotator. The 
numerical studies
presented so far have been carried on for the most irrational pair
$A$, $B$. We have found similar results for other irrationally
related pairs. We are so lead to conclude that the nature of the
observed unbounded diffusion lies in the almost-periodic
arrangement of in-kicks intervals $A$ and $B$.

To substantiate this hypothesis we have replaced the phases
$\frac{T}{2} n^2$ by two sequences of equally distributed  pseudo-random
numbers, one for the operator $U(k,A)$ and one for
its companion $U(k,B)$.  In so doing, we have obtained
quite similar results as those reported above. In
fig. \ref{betak} the exponents $\beta_{av}$ for this case are
also reported.

\section{Conclusions}

We have studied the dynamics of a quantum rotator kicked at
times generated by the almost periodic Fibonacci sequence.
Contrary to that of the usual kicked rotator,
this dynamics  nevers shows quantum localization.
We have introduced and computed various indicators of the
spreading of an initially localized wave packet, which
have permitted to show that the dynamics features regular
diffusion for large values of the kick amplitude ($k>k_c$)
and anomalous
sub-diffusion for small values $(k<k_c)$. In this latter range,
the exponent of this diffusion
appears to follow a power-law behaviour with the kick amplitude,
$\beta_{av}(k) \sim k^{\frac{2}{3}}$.
This relation, although only numerically established,
is quite interesting, and deserves in our opinion
further investigation.

Similarities and differences between the
Fibonacci kicked rotator and other quasi-periodic models
are to be noted.
We have already observed that anomalous diffusion is typically found
in almost periodic, one-dimensional lattice systems. It also appears
in the kicked Harper model. 
To the contrary,
the Fibonacci kicked rotator is qualitatively
different from the rotator acted upon by equally spaced kicks, 
with an amplitude $k$
which is a  quasi-periodic function of time.
The case when this function contains $m$
incommensurate frequencies has been studied in 
\cite{dls83,cgs89,bs97}. It was shown that this model
is a dynamical analog of the Anderson localization in a space of
effective dimension $d=m+1$. In this way the usual
kicked  rotator ($m=0$) corresponds to $d=1$,
and--of course--is always localized.
For $m=1$ the excitation is still always localized,
but the localization length grows exponentially with $k$
\cite{dls83}, in analogy with the
Anderson localization in $d=2$. Finally, for $m>1$
{\em i.e.}  $d>2$,
a transition from localization to diffusive 
excitation takes place above some critical kick
amplitude, in analogy with
the Anderson transition for $d>2$ \cite{cgs89,bs97}. 
As we see this behaviour drastically
differs from the one we find in this paper for the Fibonacci kicked rotator.

In our opinion this difference might be due to the fact
that in \cite{dls83,cgs89,bs97} the kick amplitude $k$
is an analytic function of incommensurate phases (frequencies).
For this, such a model can be mapped into an effective
solid-state hopping model, similar to the Anderson model,
with hopping only to a finite number of nearby sites.
On the contrary,
the almost-periodic sequence of unitary operators
of the Fibonacci kicked rotator
renders the situation much richer, and gives rise to
a transition from regular to anomalous diffusion.

As a matter of facts, the dynamics of the Fibonacci kicked rotator seems
more similar to a wave spreading on a two-dimensional
quasi-crystal lattice. Indeed, studies of quantum diffusion 
over an octagonal quasiperiodic tiling have shown
a similar transition from anomalous to regular diffusion
\cite{sire}. However, in spite of this first glance similarity
more detailed studies are required to establish
a quantitative relation between these models, and to
gain a better theoretical understanding of 
the results presented in this paper.

\newpage

\begin{thebibliography}{999}
\bibitem{infto} Of course, numerical experiments are not a proof that
anomalous diffusion holds on indefinitely. Yet, the large time-scale
over which 
this phenomenon is observed in fig. 3 renders this conclusion plausible.
\end{thebibliography}
\end{document}